\documentclass[a4paper,12pt]{article}

\usepackage{amsmath,amssymb,array,amsfonts}
\usepackage[dvipdfmx]{graphicx}
\usepackage{comment,xcolor}

\allowdisplaybreaks

\makeatletter
    
    \@addtoreset{equation}{section}
\makeatother

\usepackage[dvipdfmx]{hyperref} 
\hypersetup{
			linkcolor= purple,%
			colorlinks= true,
			citecolor = cyan
			}

\def\e {{\rm e}}	

\def\d {{\rm d}}

\begin{document}
\setlength{\baselineskip}{18pt}
\begin{titlepage}

\begin{flushright}
SU-HET-02-2013
\end{flushright}
\vspace{1.0cm}
\begin{center}
{\Large\bf 
$\mu$ term and supersymmetry breaking from six dimensional theory
} 
\end{center}
\vspace{25mm}

\begin{center}
{\large
Yuki Adachi 
\footnote{e-mail : y-adachi@matsue-ct.ac.jp},
Naoyuki Haba$^*$
\footnote{e-mail : haba@riko.shimane-u.ac.jp},
Toshifumi Yamashita$^{**}$
\footnote{e-mail : tyamashi@aichi-med-u.ac.jp}
}
\end{center}
\vspace{1cm}
\centerline{{\it
Department of Sciences, Matsue College of Technology,
Matsue 690-8518, Japan}}

\centerline{{\it
$^*$Graduate School of Science and Engineering, Shimane University,
Matsue 690-8504, Japan
}}

\centerline{{\it
$^{**}$ Department of Physics, Aichi Medical University, Nagakute, 
480-1195, Japan}}
%
%
\vspace{0.5cm}

\begin{abstract}
We propose  a new next-to-minimal supersymmetric standard model (NMSSM) which is on a six-dimensional spacetime compactified on a $T^2/Z_3$ orbifold.
In this model, three  gauge singlet fields $N,\ S_1$ and $S_2$ in addition to the minimal supersymmetric standard model (MSSM) fields are introduced. 
These fields are localized at some fixed points except for the singlet $N$ and the gauge fields.
The $\mu $ parameter is provided from the vacuum expectation value (vev) of $N$.
The $F$ terms get vevs simultaneously, and the gauginos mediate the supersymmetry breaking to the MSSM sector. 
Both of these parameters are strongly suppressed due to the profile of $N$.
Thus these parameters induced  from those of the order of the so-called GUT scale 
can become close to the electroweak scale without unnatural fine tuning.

\end{abstract}

\end{titlepage}




\newpage
\section{Introduction}
The standard model  seems to be established by
explaining  various physical observables especially electroweak precision measurements
and 
the discovery of the Higgs boson at the Large Hadron Collider experiment 
\cite{Chatrchyan:2012ufa}.
Though it seems to succeed, 
the Higgs boson mass is unstable under the large quantum corrections of the order of the so-called GUT scale $M_\text{GUT}$ or the Planck scale $M_\text{Pl}$.
It is known as a \lq\lq hierarchy problem\rq\rq.

Supersymmetry is a symmetry between fermions and bosons and it makes the Higgs mass stable by the cancellations of the radiative corrections away them.
It is therefore  one of the most attractive ideas solving the hierarchy problem. 
The standard model can be extended to be supersymmetric one,
{\it i.e.} the minimal supersymmetric standard model (MSSM),
by adding a superpartner of each standard model particle. 
There must be two different types of mass parameters in the MSSM
for phenomenologically acceptable model.
One is the $\mu$ parameter  and the others are soft supersymmetry breaking parameters.
Both parameters should take the order of the electroweak scale to be sufficient to correct electroweak symmetry breaking
and thus they are much smaller than the cutoff scale.
The $\mu$ parameter is expected to be generated at the cutoff scale of the MSSM around the $M_\text{GUT}$ or $M_\text{Pl}$ 
so that it naively might be much larger than the weak scale.
To avoid the above problem, 
the next-to-minimal supersymmetric standard model (NMSSM) is often considered \cite{Nilles:1982dy}.
It contains singlet chiral superfield $N$ in addition to the MSSM fields and has the term $\lambda N H_u H_d$ 
where $H_u$ and $H_d$ stand for the up- and down-type Higgs doublets respectively and 
 $\lambda$ is a dimensionless parameter
instead of the $\mu$ term.
Then, the $\mu$ term is given by the non-vanishing vacuum expectation value (vev) of the singlet field $N$. 

The soft supersymmetry breaking parameters, in contrast with the $\mu$ parameter, 
 include various kinds of parameters 
such as mass terms of sparticles, trilinear couplings and so on.
Their scale should be also around the weak scale for the hierarchy problem to be solved.
Then, their pattern is highly restricted by the experiments, especially the flavor changing neutral current (FCNC) processes, and the regions of the parameter space is constrained, 
{\it e.g.} 
the squarks and sleptons masses are nearly degenerate.
%
To realize this, one simple way is to have a compact extra dimension with a radius around $M_\text{GUT}$
where the gauge supermultiplets of the MSSM can propagate in the bulk.
The matter fields of the MSSM  such as the quarks and the Higgs fields are bounded at a certain fixed point and the source of supersymmetry breaking is put at another fixed point.
Only the gauge supermultiplets communicate to the supersymmetry breaking sector,  
and the gauginos become massive  by local interactions at the supersymmetry breaking sector.
After integrating out the extra dimension, four-dimensional MSSM  including nonzero gaugino masses are obtained at the GUT scale.
Then the squarks and sleptons get  masses from the massive gauginos through the renormalization group evolution.
The large FCNC processes vanishes because  the Yukawa couplings are the only source of the flavor violation.
There are only gaugino mass parameters and all of the other soft supersymmetry breaking parameters are determined by it
so that this scenario is very predictable. 
In this type of scenarios, the gauginos behave as the messenger and it is called \lq\lq gaugino mediation \rq\rq  \cite{GM1}.

In this paper, we propose a simple NMSSM model
 which generates suitable $\mu$ term and supersymmetry breaking from the GUT scale in the context of an extra-dimension scenario.
We provide the model in the next section and discuss how the small  parameters are achieved.
We summarize this model in section 3.
The effects of the KK modes are discussed in Appendix~\ref{sectionKKmode}.

\section{Model}
\begin{figure}[h]
\begin{center}
\includegraphics[scale=1.2]{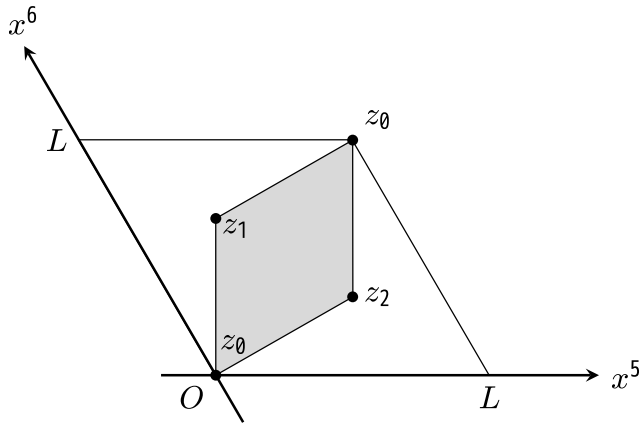}

\caption{$T^2/Z_3$ compactification: fundamental domain and three fixed points are depicted.}
\label{figure1}
\end{center}
\end{figure}

In this section, we provide the model.
We consider a six-dimensional spacetime compactified on a $T^2/Z_3$ orbifold
with the circumference $L$
\cite{Dixon:1985jw,T2Z31,T2Z32}
.
The extra dimensions are labeled in $(x^5,\ x^6)$ and 
the angle between them is $\frac{2\pi}{3}$ to be compatible with a $Z_3$ action.
On this compactification, the point $x^5+L\ (x^6+L)$ is identified  as $x^5 \ (x^6)$.
The unit vectors $\vec e_5$ and $\vec e_6$ along with the axes $x^5$ and $x^6$ are defined by 
\begin{equation}
|\vec e_5|=|\vec e_6|=1,\ 
\vec e_5\cdot \vec e_6=\cos \frac{2\pi}{3}=-\frac{1}{2}.
\end{equation}
Then the three fixed points appear at
\begin{equation}
z_0=0,\ 
z_1=\frac{L}{3}(\vec e_5+2\vec e_6),\ 
z_2=\frac{L}{3}(2\vec e_5+\vec e_6),
\end{equation}
as is shown in Figure \ref{figure1}. 
In this setup, the four-dimensional $\mathcal N = 2$ supersymmetry is reduced to $\mathcal N=1$ by the $Z_3$ orbifolding
\cite{Dixon:1985jw,T2Z31}
.

Before describing our model precisely, we give  some properties specific to the compactified six-dimensional theory.

The six-dimensional gauge coupling constant $g_{\rm 6D}$ carries mass dimension $-1$.  
Then the higher-dimensional operators appear in loop corrections with the effective coupling
($g_{\rm 6D}M$)~\cite{GM1}, where $M$ is the cutoff scale. 
If the effective coupling is smaller than $4\pi$ , 
this effective theory is perturbative and predictive. On the
other hand, after the dimensional reduction, the four-dimensional
gauge coupling constant $g$ becomes
$g=\frac{g_{\rm \rm 6D}}{\sqrt{\frac{\sqrt{3}}{2}}L}$.
Combining with the above conditions, a relation between the parameters $g,\ L$ and $M$ are obtained by
\begin{equation}
 1 
\geq  
 \frac{(g_{\rm 6D} M)^2}{(4\pi)^2}
=
 \frac{1}{(4\pi)^2}\left(\frac{g_{\rm 6D}}{\sqrt{\frac{\sqrt{3}}{2}}L}\right)^2
 \frac{\sqrt{3}}{2}(LM)^2
=
 \left(\frac{g}{4\pi}\right)^2\cdot \frac{\sqrt{3}}{2}(LM)^2.
\end{equation}
We require the four-dimensional gauge coupling constant is perturbative as
$\alpha=g^2/4\pi\sim \mathcal O(10^{-2})$,
it should be
\begin{equation}
LM\leq \sqrt{\frac{2}{\sqrt{3}}}\frac{4\pi}{g}\sim \mathcal O(10).
\end{equation}
To reproduce the unified gauge coupling constant at the GUT scale,
we must choose $LM \leq 19$.
Thus, we assume that the compactification scale $L$ is worth ten percent of the cutoff scale of the six-dimensional theory through this paper. 

The six-dimensional Planck scale $M_6$ is given by
\begin{equation}
 M_6
=
 \sqrt{\frac{M_{\rm Pl}}{L}}
=
 \mathcal O(10^{17}{\rm GeV}),
\end{equation}
where we input $L\sim M_{\rm GUT}$.
If we assume the six-dimensional Planck scale $M_6$ as the cutoff $M$ of this six-dimensional theory, 
we expect to construct the reasonable model.

\subsection{Lagrangian}
Let us show our model.
We introduce  three singlet fields $N,\ S_1$ and $S_2$ in addition to the MSSM fields.
The MSSM fields live at the fixed point $z_0$ except for  the gauge supermultiplets which propagate the extra dimensions.
The singlets $S_1$ and  $S_2$ are put at the fixed points $z_1$ and $z_2$, respectively,
and the singlet $N$ has a Gaussian profile localized around $z_1$ as follows:
\begin{subequations}
\begin{align}
&
N(x^I)=N(x^\mu) M f(x^5,x^6),
\\
\label{profile}
&
f(x^5,x^6)= c\exp\left[-M^2 \left\{(x^5-L/3)\vec e_5 +(x^6-2L/3)\vec e_6\right\}^2\right],
\end{align}
where $I=0,1,2,3,5,6$ and $\mu=0,1,2,3$.
The normalization factor $c$ is defined by 
\begin{equation}
c^{-2}=M^2\int_{0}^L  \d  x^5 \int_{0}^L \d x^6\frac{\sqrt{3}}{2} |f\left (x^5,x^6\right)|^2.
\end{equation}
\end{subequations}
For the sufficiently large $ML$ \cite{ArkaniHamed:1999dc}, the profile $f(x^5,x^6)$ is extremely small at $z_0,\ z_2$, namely, 
\begin{equation}
f(z_1)=c,\ f(z_0)=f(z_2)=c\e^{-M^2L^2/3}=c\epsilon\ll c\ . 
\end{equation}

The Lagrangian in this scenario is described by three parts: 
\begin{equation}
 \mathcal L
=
 \mathcal L_\text{gauge} +\mathcal L_\text{MSSM matter} + \mathcal L_\text{singlets}
 .
\end{equation}
The first part $\mathcal L_\text{gauge}$ includes the MSSM gauge supermultiplets. 
After integrating out the $x^5$ and $x^6$, the (zero mode) massless gauge bosons and gauginos appear, whose Lagrangian is given as 
\begin{align}
 \mathcal L_\text{gauge} 
=&
 \iint \d x^5 \d x^6\frac{\sqrt{3}}{2} \mathcal L_\text{gauge}^\text{6D}\notag\\
=&
 \iint \d x^5 \d x^6\frac{\sqrt{3}}{2} 
 \left[ 
  -\frac{1}{2}{\rm Tr} \left(F_{IJ} F^{IJ}\right) 
  +{\rm Tr} (\bar \lambda i\Gamma^I D_I \lambda) +\cdots
 \right],
\end{align}
where $I,J=0,1,2,3,5,6$.
Thus the zero mode gauginos $\lambda$ have normalization factor 
$\sqrt{2/(\sqrt{3}L^2)}$.
The MSSM matter fields which are localized at the fixed point $z_0$ are described by $\mathcal L_\text{MSSM matter}$.
 
The Lagrangian of the singlet fields  $\mathcal L_\text{singlet}$ includes the kinetic terms and interactions.
We impose a $U(1)_R$ symmetry on this model with the following charge assignment.
\begin{equation}
Q(H_u)=Q(H_d)=Q(N)=\frac{2}{3},\ 
Q(S_1)=Q(S_2)=\frac{4}{3}.
\end{equation}
We assume that $U(1)_R$ is explicitly broken at the fixed point $z_1$ by mass
parameters that have positive $U(1)_R$ charges.
These mass parameters are expected to be vevs of some fields in an  underlying theory.
Then the allowed interactions become 
\begin{align}
 W
=& 
 \iint \d x^5\d x^6 \frac{\sqrt{3}}{2}
 \left[
 \left[\kappa \frac{N(x^I)}{M} H_u H_d +a_0 \left(\frac{N(x^I)}{M}\right)^3\right]_{z_0}
 +\left[m_2 S_2\frac{N(x^I)}{M}+a_2 \left(\frac{N(x^I)}{M}\right)^3\right]_{z_2}
 \right.
 \notag\\
 &\hspace{40pt}
\left.
 +\left[m_1 S_1\left(\frac{N(x^I)}{M}-v\right) + a_1\left(\frac{N(x^I)}{M}\right)^3 +m_3\left(\frac{N(x^I)}{M}\right)^2+m_4^2 \frac{N(x^I)}{M} \right]_{z_1}
\right]
\notag
 \\
=&
 \kappa \epsilon c NH_uH_d+m_2 \epsilon c S_2N+
  m_1S_1(cN-v) + (a_0\epsilon^3+a_1+a_2\epsilon^3)c^3 N^3 +c^2m_3 N^2 +cm_4^2N,
\end{align}
where $H_u$ and $H_d$ stand for up- and down-type Higgs doublets, respectively.
The mass parameters $m_1,\ m_2,\ m_3,\ m_4$ and $v$ are an order of the cutoff scale in this model and $\kappa ,a_0,a_1$ and $a_2$ are dimensionless parameters.
Note that the Higgsino mass parameter $\mu$ which would be as large as the cutoff scale is forbidden by the $U(1)_R$ symmetry.
An important point of this superpotential is that the original mass parameters are the high energy scale such as the GUT scale, 
but the some of them are strongly suppressed due to the Gaussian profile of $N$ at the low-energy effective theory.
Then we expect that the singlet $N$ takes a large vev  
but it produces a desirable $\mu$ parameter.
Here we comment on effects of KK modes of the singlet $N$.
Since their profiles are different from the one of the zero mode and are not necessarily 
 suppressed at the fixed points, 
 they might induce large $\mu$ term and/or tadpole term of the singlet $S_2$,
which change the above superpotential to break our framework.
Actually, they are expected to be strongly suppressed similar to the
zero mode, as 
argued in appendix \ref{sectionKKmode}.

Let us investigate the vacuum of this potential. 
We ignore the term $N,N^2$ and $N^3$  since they just shift the vevs of the singlets $S_1$ and $S_2$.
The $F$ terms become 
\begin{subequations}
\begin{align}
&F_{S_1}=m_1 \left( c N-v\right) ,
\\
&F_{S_2}=c\epsilon m_2 N, \\
&F_{N}=c\epsilon m_2 S_2+c m_1 S_1+\kappa c\epsilon H_d H_u ,
\\
&F_{H_u}=\kappa c\epsilon H_d N,
\\
&F_{H_d}=\kappa c\epsilon H_u N.
\end{align}
\end{subequations}
The scalar potential which is defined by $V=\sum_i |F_i|^2$  is
\begin{equation}
 V
=
 c^2\left( \epsilon m_2 S_2+m_1 S_1+\kappa\epsilon H_d H_u \right) ^2
 +{m_1}^{2} {\left( c N-v\right) }^{2}
 +c^2\epsilon^2 {m_2}^{2} {N}^{2}
 +c^2\epsilon^2 {\kappa}^{2} N^2({H_u}^{2} +{H_d}^2).
\end{equation}
The vacuum will be given by minimizing above potential, 
and we obtain 
\begin{subequations}
\begin{align}
\langle S_1\rangle =- \frac{\epsilon m_2 C}{m_1} ,\ 
\langle S_2\rangle =C,\ 
\langle N\rangle =\frac{ 5{m_1}^{2} v}{{c(4\epsilon}^{2} {m_2}^{2}+ {5m_1}^{2})},\ 
\langle H_u\rangle = 
\langle H_d\rangle =0, 
\end{align}
\end{subequations}
where $C$ is an arbitrary constant.
Some $F$ terms get non-vanishing vevs as follows:
\begin{subequations}
\begin{align}
&\langle F_{S_1}\rangle=-\frac{4{\epsilon}^{2} m_1 {m_2}^{2} v}{4{\epsilon}^{2} {m_2}^{2}+5{m_1}^{2}},\ 
\langle F_{S_2}\rangle=\frac{5 \epsilon {m_1}^{2} m_2 v}{4{\epsilon}^{2} {m_2}^{2}+5{m_1}^{2}}
,\\
&\langle F_N\rangle=\langle F_{H_u}\rangle =\langle F_{H_d}\rangle =0.
\end{align}
\end{subequations}
As we mentioned in the above section, the factor $\epsilon$ is small by the locality of $N$,
and the derived $F$ terms arise a suitable supersymmetry breaking scale.

\paragraph{$\mu$ parameter:}
The Higgsino mass parameter $\mu$ in this model is achieved by the vev $\langle N\rangle$ as we see in the above paragraphs:
\begin{equation}
\mu =\kappa \epsilon c \langle N\rangle = \kappa \epsilon v +\mathcal O(\epsilon^2).
\end{equation}
Though the mass parameter $v$ is an order of $M_{\rm GUT}$ scale, 
the exponential factor $\epsilon = \e^{-(ML)^2/3}$ appears and the $\mu$ parameter is strongly suppressed.
Thus the desirable $\mu$ parameter is realized by order one tuning of $ML$.

Note that this potential has a flat direction and massless particles will appear
\footnote{
We may assume the flat direction is stabilized by a certain effect. 
For instance, the higher-dimensional terms  $|S_2|^4/M^2-|S_2|^6/M^4$ in the K\"ahler potential fixes $C=\pm M/\sqrt2$.
}
. 
However, there is no direct communication between these massless fields $S_1,S_2$ and the MSSM fields 
since they are separated into different fixed points.
The bulk gauge fields and the singlet $N$ could connect the massless fields and the 
 MSSM fields, since they have overlap on the both fixed points. 
In fact, however, they have no or very weak interactions with the massless fields and the MSSM fields ($H_u,H_d$), respectively.
Thus, these massless fields have little effect on phenomenology in the MSSM sector
\footnote{
These massless fields may raise the so-called Polonyi problem
\cite{Coughlan:1983ci,Ellis:1990nb}
.
In this case,
a certain modification is needed
\cite{Randall:1994fr,  Banks:1995dt}
.
}
.

\subsection{Mediating supersymmetry breaking}
In this section, we discuss  how the supersymmetry breaking is brought to the MSSM sector.
Since the MSSM gauge supermultiplets live on the bulk, the  gauginos can mediate the supersymmetry breaking from the invisible sector to the MSSM sector.
It is known as a \lq\lq gaugino mediation\rq\rq \cite{GM1}
, {\it i.e.}, gauginos behave as messengers.
In our model, the nonzero $F$ term at the fixed point $z_2$ generates 
the localized gaugino mass terms, for example, through the gauge mediation.
After integrating out the extra dimensions, 
the gauginos obtain the following masses,
\begin{equation}
 m_\lambda
\sim
 \frac{\alpha_0}{4\pi}\frac{\langle F_{S_2}\rangle}{\langle S_2 \rangle}\frac{1}{M^2}\left(\frac{1}{\sqrt{\frac{2}{\sqrt{3}}}L}\right)^2
=
 \frac{\alpha_0}{4\pi}\frac{\langle F_{S_2}\rangle}{\langle S_2\rangle}\frac{\sqrt{3}}{2 M^2L^2}
=\frac{\sqrt{6}\alpha_0}{8\pi}\frac{1}{(ML)^2}\frac{\mu}{\kappa},
\end{equation}
from the local mass terms  at $z_2$.
The $\alpha_0$ stands for the fine structure constant at the GUT scale.
Note that $S_1$ has also a nonzero $F$ term but it is negligible 
and we ignore the contribution from $S_1$.

The other soft supersymmetry breaking parameters are generated by the 
four-dimensional renormalization group evolution below the compactification scale
.
It corresponds to a boundary condition that the only gaugino mass $m_\lambda$ is given at the input scale,
so this mechanism is similar to the \lq\lq no-scale \rq\rq type of the Planck-mediated supersymmetry breaking
\cite{noscale}
.
The supersymmetry breaking parameters are controlled by the gaugino mass $m_\lambda$ at the GUT scale
and the Yukawa couplings are the only source of the flavor violation so that the dangerous FCNC processes are expected to be suppressed.

Let us explore the parameter space of this model.
For the case $ML\sim 9$ and $\kappa\sim 10^{-4}$, we obtain the following proper values,
\begin{equation}
m_\lambda 
  \sim \frac{\sqrt{6}\alpha_0}{8\pi}\frac{1}{(ML)^2}\epsilon M
  \sim \mathcal O(\text{TeV}),~~
\mu
  \sim \frac{8\pi}{\sqrt{6}\alpha_0}(ML)^2 \kappa m_\lambda
  \sim \mathcal O (\text{TeV}),
\end{equation}
where we use $\alpha_0\sim 1/24$.
One can introduce another dimensionless parameter $k$ to the profile $f(x^5,x^6)$ in 
eq. (\ref{profile}) as follows:
\begin{equation}
f(x^5,x^6)= c\exp\left[-k^2M^2 \left\{(x^5-L/3)\vec e_5 +(x^6-2L/3)\vec e_6\right\}^2\right].
\end{equation}
As a result of this, the derived gaugino mass $m_\lambda$ and  $\mu$ parameter at the cutoff scale are given by replacing $\e^{-(ML)^2/3}$ to $\e^{-(kML)^2/3}$;
\begin{equation}
m_\lambda 
  \sim \e^{-(kML)^2/3}\frac{\sqrt{6}\alpha_0}{8\pi}\frac{1}{(ML)^2} M
  ,~~
\mu
  \sim \frac{8\pi}{\sqrt{6}\alpha_0}(ML)^2 \kappa m_\lambda
  .
\end{equation}
which allows us to choose the  milder parameters as follows:
\begin{equation}
k\sim 7,\  \frac{1}{\kappa}\sim 440,\  M\sim 1.3M_{\rm GUT}.
\end{equation}

We finally comment on the Higgs mass in brief.
Recent discovery of the Higgs boson around 126~GeV puts strong constraints on the MSSM parameter space.
It suggests that the large $A_t$ term is desirable to lift up the  Higgs boson mass through the stop loop corrections
\cite{Aterm}.
In this model, however, the origin of the supersymmetry breaking parameter is 
the gaugino mass $m_\lambda$
so that the  $A_t$ term is relatively small compared with the required one,
{\it i.e} there is no free parameter to tune the $A_t$ term
\footnote{We also note that this model reduces to the MSSM in the low energy effective theory and  
the singlet $N$ has no effects on the Higgs boson mass.}.
It is a common problem to the models that have the no-scale type boundary condition.
In such cases,  the Higgs boson mass is less than  123~GeV 
\cite{higgsmass_noscale}.

There are several way to push up the Higgs boson mass in our model.
Examples are adding vector-like matters; extra gauge fields; additional singlets and/or $SU(2)$ triplets.
The first is introducing the (s)top-like matter
to enhance the quantum corrections to the Higgs boson mass \cite{vector_like}. 
The second is a way to bring additional source for the quartic coupling of the Higgs doublets by the additive $D$ terms and increase the tree level Higgs boson mass \cite{extraDterm}.
The last is a way to bring additional source for the quartic coupling of the Higgs doublets by the additive $F$ terms~\cite{Nilles:1982dy,Espinosa,gGHU}.

\section{Summary}
In this paper we proposed the model realizing the TeV scale supersymmetry breaking parameters and the $\mu$ parameter.
In this model, the six-dimensional space-time compactified on the $T^2/Z_3$ is considered,
and the radius of the torus is set by the GUT scale. 
There are three fixed points $z_0,z_1$ and $z_2$ on the compactified space,
and we put the MSSM matter fields at the fixed point $z_0$ though the  MSSM gauge fields can propagate the extra dimensions.
We also introduce the three gauge singlet fields $S_1,S_2$ and $N$ in addition to the MSSM fields.
$S_1$ and $S_2$ are put at the fixed points $z_1$ and $z_2$ respectively, 
and $N$ is localized around the fixed point $z_1$ which has the Gaussian profile.

We introduce the NMSSM-like superpotential. 
Though the superpotential have the $\mathcal O(1)$ coupling between the MSSM Higgs doublets and $N$ above the compactification scale, 
it is strongly suppressed below the compactification scale due to the locality of $N$.
After minimizing the scalar potential, the singlet $N$ takes vev and non-zero $F$ terms  of $S_1$ and $S_2$ appear at the fixed points $z_1$ and $z_2$, simultaneously.

Then the gauginos obtain  masses from these non-zero $F$ terms at these fixed points,
and the MSSM matter fields at the fixed point $z_0$ receive the supersymmetry breaking by these massive gauginos.
Namely, the  gauginos behave as messengers (gaugino mediation).
The other soft mass terms are induced by the four-dimensional renormalization group evolution
and 
the dangerous FCNC processes are suppressed.
The Higgsino mass parameter $\mu$ is generated by the vev of the singlet $N$. 
These are strongly suppressed by the wave function of $N$ and thus the electroweak 
scale can be achieved from parameters of order of the GUT scale by order one tuning. 
We explore the parameter space and  obtain a suitable result which realizes desirable supersymmetry breaking and $\mu$ parameter.

\subsection*{Acknowledgments} 

This work is partially supported by Scientific Grant by Ministry of
Education and Science, Nos. 24540272,  23340070,  21244036, 22011005,
and by the SUHARA Memorial Foundation.


\appendix

\section{Effects of the KK modes}
\label{sectionKKmode}

In this appendix, we will give a short discussion on the effects 
 caused by the nonzero KK modes of the singlet $N$,
 which are not argued precisely in the main text. 
The KK modes of the singlet $N$ are not necessarily localize as the zero mode,
 so that they might induce large $\mu$ term and/or tadpole term of $S_2$ which 
 destroy our framework.

From now on, we investigate these effects by use of 
 higher-dimensional propagator~\cite{5Dprop} of $N$.
We consider five-dimensional setup, for simplicity,   
 where the fifth dimension $x^5$ is compactified.
Note that the derived results are expected to be similar in six dimensions, 
as the Yukawa suppression $\e^{-mr}$ is common in any dimensions.
To realize the Gaussian profile $\e^{-\frac{1}{2}(mx^5)^2}$, 
we assume the singlet $N$ satisfies the following equation
\begin{equation}
\label{A1}
\left[\partial_\mu \partial^\mu  - (-\partial_5 +m^2 x^5)(\partial_5 + m^2x^5)\right] N(x^\mu,x^5)=0,
\end{equation}
and 
it obeys the boundary conditions 
\begin{equation}
\label{boundaryconditions}
\partial_5 N|_{x^5=0} = 0,\, (\partial_5  +m^2L)N|_{x^5=L}=0,
\end{equation}
where $L$ stands for the compactification scale.
The equation (\ref{A1}) is decomposed into two parts;
One is an ordinary four-momentum relation $\partial _\mu \partial^\mu N =-p_\mu p^\mu N$ and the other is
\begin{equation}
\label{EOM}
\left[-p^2  - (-\partial_5 +m^2 x^5)(\partial_5 + m^2x^5)\right] f(p^2,x^5)=0,
\end{equation}
where $f(p^2,x^5)$ stands for the mode functions of $N(x^\mu,x^5)$.
The propagation along with the extra dimension from $x^5$ to $x'^5$ accompanied with momentum $p$ is described by a propagator $P(p^2,x^5,x'^5)$
defined by
\begin{equation}
\left[-p^2  - (-\partial_5 +m^2 x^5)(\partial_5 + m^2x^5)\right]
P(p^2,x^5,x'^5) = -\delta(x^5-x'^5).
\end{equation}
Note that the above propagator $P(p^2,x^5,x'^5)$ includes all the effects of the KK modes.
It is understood if we expand the delta function in the series $f_{M_{KK}}$ which satisfies 
$(-\partial_5+m^2 x^5)(\partial_5 + m^2 x^5) f_{M_{KK}}=M_{KK}^2f_{M_{KK}}$.
The above propagator is then described as
\begin{equation}
P(p^2,x^5,x'^5) = \sum_{M_{KK}} \frac{f_{M_{KK}}(x^5) f_{M_{KK}}(x'^5)}{p^2 +M_{KK}^2}.
\end{equation}
The mode function $f_{M_{KK}}(x^5)$ and $f_{M_{KK}}(x'^5)$ are  proportional to the couplings between
the KK mode of the singlet $N$ and the other fields at the $x^5$ and $x'^5$, respectively.
This means that the propagator includes all the effects of KK modes propagating 
 from $x^5$ to $x'^5$.

Next we derive the propagator  $P(p^2 ,x^5,x'^5)$.
It satisfies the equation (\ref{EOM}) except for the $x^5\neq x'^5$
so that it can be constructed as $f(p^2,x^5)f(p^2,x'^5)$.
The mode function $f(p^2,x^5)$ is divided into an even function $f_A$ and an odd one $f_B$ at $x^5=0$ as
\begin{equation}
f(p^2,x^5)=A f_A(p^2,x^5) + Bf_B(p^2,x^5),
\end{equation}
where $A$ and $B$ are arbitrary constants.
Then the propagator which satisfies the boundary condition (\ref{boundaryconditions}) becomes
\begin{equation}
\label{propagator}
P(p^2,x^5,x'^5)=f_A(p^2,x^5)[A'f_A(p^2,x'^5)+B'f_B(p^2,x'^5)],
\end{equation}
with 
\begin{equation}
\label{cond1}
-\frac{A'}{B'} 
 =\left.
  \frac{(\partial_5+m^2L)f_B(p^2,x^5)}{(\partial_5+m^2L)f_A(p^2,x^5)}
  \right|_{x^5=L}.
\end{equation}
Since the derived propagator (\ref{propagator}) also satisfies the equation (\ref{EOM}) at $x^5=x'^5$,
we have
\begin{equation}
\label{cond2}
-1
  =B'
  \left[
  \partial'_5 f_A(p^2,x'^5) f_B(p^2,x'^5) -\partial'_5 f_B(p^2,x'^5)f_A(p^2,x'^5)
  \right].
\end{equation}
These constants $A'$ and $B'$ are determined by the conditions (\ref{cond1}) and (\ref{cond2}).

Since we are interested in whether the KK modes have large contributions
 below the compactification scale or not,  
 we concentrate on the case of low momentum $p^2\to 0$.
The mode functions can be expanded in terms of the momentum $p^2$ as 
\begin{equation}
f_A(p^2,x^5)= f_A^0 +p^2 f_A^1+p^4 f_A^2+\cdots . 
\end{equation}
They are obtained by solving the equation (\ref{EOM}) in a perturbative manner 
and we have
\begin{align}
&f_A^0(x^5)=\e^{-\frac{1}{2}(mx^5)^2} ,
\\
&f_B^0(x^5)=\e^{-\frac{1}{2}(mx^5)^2}\int_0^{x^5}\d x  \e^{m^2x^2} , 
\\
&f_A^1(x^5) = \e^{-\frac{1}{2}(mx^5)^2}\int_0^{x^5} \d x \e^{m^2x^2}\int_0^x \d t \e^{-\frac{1}{2}m^2 t^2} f_A^0(t),
\\
&f_B^1(x^5) = \e^{-\frac{1}{2}(mx^5)^2}\int_0^{x^5} \d x \e^{m^2x^2}\int_0^x \d t \e^{-\frac{1}{2}m^2 t^2} f_B^0(t),
\\
&f_A^2(x^5) = \e^{-\frac{1}{2}(mx^5)^2}\int_0^{x^5} \d x \e^{m^2x^2}\int_0^x \d t \e^{-\frac{1}{2}m^2 t^2} f_A^1(t),
\\
&f_B^2(x^5) = \e^{-\frac{1}{2}(mx^5)^2}\int_0^{x^5} \d x \e^{m^2x^2}\int_0^x \d t \e^{-\frac{1}{2}m^2 t^2} f_B^1(t).
\end{align}
Then the propagator is approximated as 
\begin{align}
P(p^2,x^5,x'^5)
 =&
 \label{derived_propagator}
  \frac{1}{ap^2}f_A^0(x^5)f_A^0(x'^5)
  \notag
 \\
 +&\frac{1}{a}
 \left[
 (c-b/a)f_A^0(x^5)f_A^0(x'^5)+f_A^0(x^5)f_A^1(x'^5)+f_A^1(x^5)f_A^0(x'^5)
 \right]
 -f_A^0(x^5)f_B^0(x'^5)
 +\mathcal O(p^2),
\end{align}
where
\begin{equation}
a=\int_0^L\d t \e^{-\frac{1}{2}m^2t^2}f_A^0(t),\,
b=\int_0^L\d t \e^{-\frac{1}{2}m^2t^2}f_A^1(t),\,
c=\int_0^L\d t \e^{-\frac{1}{2}m^2t^2}f_B^0(t).
\end{equation}
The first line in (\ref{derived_propagator}) corresponds to the contributions
of the zero mode which reflects the fact that the zero mode is massless. 
The second line corresponds to the contributions of the KK modes.

To estimate these effects of KK modes, we set $x^5=0$ and $x'^5=L$ 
and then the second line in (\ref{derived_propagator}) becomes
\begin{equation}
 \frac{\e^{-\frac{(mL)^2}{2}}}{\int_0^L \d t \e^{-m^2t^2}}
 \left[
 \frac{1}{\int_0^L \d t \e^{-m^2t^2}}
  \int_0^L \d z \e^{-m^2z^2}\int_0^z \d y \int_y^L \d t \e^{-m^2(t^2-y^2)}
 -\int_0^L \d y \int_y^L \d t \e^{-m^2(t^2-y^2)}
 \right].
\end{equation}
The integrations in the square brackets 
converge since
the $t^2-y^2$ is always positive through the integration.
It shows that the effects of the singlet $N$ are controlled by the Gaussian function similar in the contributions of the zero mode, and thus,
we conclude that they are not too large to destroy our framework.


\end{document}